# Observation of Negative Contact Resistances in Graphene Field-Effect Transistors

Ryo Nouchi,[1,a] Tatsuya Saito,[2] and Katsumi Tanigaki[1,2]

[1] *WPI-Advanced Institute for Materials Research (WPI-AIMR), Tohoku University, Sendai 980-8577, Japan*

[2] *Department of Physics, Tohoku University, Sendai 980-8578, Japan*

The gate-voltage ($V_G$) dependence of the contact resistance ($R_C$) in graphene field-effect transistors is characterized by the transmission line model. The $R_C$-$V_G$ characteristics of Ag, Cu, and Au contacts display a dip around the charge neutrality point, and become even negative with Ag contacts. The dip structure is well reproduced by a model calculation that considers a metal-contact-induced potential variation near the metal contact edges. The apparently negative $R_C$ originates with the carrier doping from the metal contacts to the graphene channel and appears when the doping effect is more substantial than the actual contact resistance precisely at the contacts. The negative $R_C$ can appear at the metal contacts to Dirac-cone systems such as graphene.

[a] Author to whom correspondence should be addressed. Present address: Nanoscience and Nanotechnology Research Center, Osaka Prefecture University, Sakai 599-8570, Japan. Electronic mail: r-nouchi@21c.osakafu-u.ac.jp.

# 1. INTRODUCTION

Charge carriers in graphene show intrinsically ultrahigh mobility,[1] and thus graphene is now recognized as a promising material for future electronic devices. The carrier transport properties should be measured using metallic electrodes. However, metal-graphene contacts introduce an additional resistance, known as contact resistance. This resistance is a limiting factor for the performance of electronic devices, and an increasing amount of research has concentrated on this issue.[2-8] The relative contribution of contact resistance to the total device resistance becomes larger in devices with shorter inter-electrode spacings, *i.e.*, in shorter channel devices. Therefore, contact resistance becomes a predominant factor to consider when attempting to achieve miniaturization and integration of graphene devices.

The four-terminal measurement or the so-called transmission line model (TLM) is commonly employed to extract and exclude the contact resistance. It has been reported that the metal-graphene contact resistance extracted by the four-terminal measurement can be unexpectedly negative in some cases.[2] Contact resistances at metal-graphene interfaces have been considered to originate from carrier tunneling processes at the time of charge injection and extraction through the interfaces,[5] or been determined by the number of conduction modes in graphene.[7] These resistances must be positive, contrary to the extracted negative value.

In this paper, we perform detailed analyses on the apparently negative contact resistances of metal-graphene contacts. Contact resistances of graphene field-effect transistors (FETs) with Ag, Cu, and Au contacts are extracted using the TLM. The gate-voltage dependency of the extracted resistances is well reproduced by a simple model[9,10] that considers the diffusive transport of charge carriers in graphene and carrier doping from metallic electrodes to the graphene channel.

# 2. EXPERIMENTAL

Graphene layers were mechanically exfoliated onto a Si substrate covered with a 300 nm thick thermal oxide layer. The graphene layers were determined to be single-layer graphene from the optical contrast and Raman scattering spectroscopy. The Si substrate was degenerately doped and used as a gate electrode. Source and drain electrodes were fabricated on the graphene layers by conventional lithographic procedures (resist patterning by electron beam lithography, thermal deposition of metals, and liftoff). Ag, Cu, and Au were used as the source/drain contact metals. A 200 nm thick layer of Au was deposited as a cap layer against oxidation after Cu or Ag deposition; a Cr layer less than 1 nm thick was deposited prior to Au deposition as an adhesion layer. The graphene channel widths $W$, of the Au-, Cu-, and Ag-contacted devices were around 2.0, 3.5, and 3.0 μm, respectively.

The constructed devices were introduced into a vacuum probe system (base pressure: ca. $10^{-2}$ Pa), and heated at 120 °C for 12 h to remove resist residues. It was confirmed that the transfer characteristics (gate-voltage $V_G$, dependence of the drain current, $I_D$) of two FETs with the same channel length on the same graphene sheet became almost identical after the heating process. Without the heating process, significant inhomogeneity remains, even on the same graphene sheet.[2] All FET measurements were performed under the same vacuum conditions after the heating procedure, with the drain-source voltage $V_{DS}$ set to 10 mV.

## 3. RESULTS

Figure 1a shows a schematic of several source and drain electrodes with different channel lengths fabricated on a single flake of graphene mechanically exfoliated onto a $SiO_2$/Si substrate. A measured two-terminal resistance, $R = V_{DS}/I_D$, can be expressed as $R = R_G + 2R_C$, where $R_G$ is the resistance of the graphene channel, and the coefficient 2 represents the contact resistances $R_C$ at two contacts (source and drain). $R_C$ extraction from the TLM assumes homogeneous sheet resistance $\rho_G$, of the channel, *i.e.*, $R_G = (L/W)\rho_G$, where $L$ and $W$ are the graphene channel length and width, respectively. Thus, as illustrated in Fig. 1b, $R_C$ can be extracted from the intercept at $L = 0$ for a linear fit of the $R$-$L$ plot.

Figure 2a shows the transfer characteristics (gate-voltage $V_{GS}$, dependence of the drain current, $I_D$) for Ag-contacted graphene FETs. The channel length $L$, was varied from 2.5 to 0.5 μm so that the contact resistance $R_C$ could be extracted using the TLM. Since the graphene flakes were used as formed in this study, the graphene channel width $W$ was not completely uniform. To correct the channel-width variation, the total conductivity including contact resistances, which is equal to $I_D L/(V_{DS} W)$, is used instead to precisely extract $R_C$. The transfer curves shown in Fig. 2a indicates a shift of the minimum conductivity point (the so-called Dirac point or charge neutrality point, $V_{NP}$), where $V_{NP}$ of the Ag-contacted devices is shifted towards more positive $V_G$ values as $L$ becomes shorter.[11] In order to know the $V_G$-dependence of $R_C$, $R$ at the same normalized gate voltage, $V_G - V_{NP}$, was plotted against $L$ in Fig. 2b. Figure 2c shows $R$-$L$ plots for the Ag-contacted devices at $V_G = V_{NP}$. The straight line is a result of a least-squares fit, and the $R$-axis intercept of the fitting line is negative, which indicates a "negative" contact resistance according to the TLM analysis.

The extracted $R_C$ values of Ag-graphene contacts are plotted in Fig. 3a against the normalized gate voltage. As previously reported for Ti/Au contacts,[2] the $R_C$ of the Ag contacts is negative around the minimum conductivity point. The $R_C$-$V_G$ characteristics for Cu and Au contacts are shown in Figs. 3b and 3c, respectively. Although the extracted $R_C$ values for Cu and Au contacts were positive for the measured $V_G$ range, they display a decrease around the minimum conductivity point, as with the Ag contacts.

The actual contact resistance at the metal-graphene interfaces originates from charge injection (and extraction) through the interfaces, and must be positive. Therefore, this result indicates that the TLM and the four-terminal measurement fail to correctly extract the metal-graphene contact resistances. The failure of these methods can be explained by the carrier doping from the metallic electrodes to the graphene channel.[11,12] At the minimum conductivity point (the charge-density profile of $V_G = 0$ in Fig. 4a), the central region of the graphene channel is almost charge neutral, while the regions close to the metal contacts receive charge carriers from the electrodes and should be more conductive than the central region; the existence of the doped regions apparently reduces $R_C$ in this case. With $V_G$'s far from the minimum conductivity point (the charge-density profiles of $V_G > 0$ and $V_G < 0$ in Fig. 4a), the central region is significantly doped by the field-effect gating, while the regions near the contacts should be less conductive than the central region, due to the $V_G$-uncontrollable charge density at the contacts (charge-density pinning)[13,14]; the doped regions apparently increases $R_C$ in this case. Consequently, $R_C$ values extracted by the TLM or four-terminal measurement include a contribution from the additional resistance due to the metal-contact doping, $R_{CD}$, in addition to the actual contact resistance precisely at the metal-graphene contacts, $R_{CI}$ (most simply, the resistance originating from a tunneling effect; Fig. 4b). If the reduction in $R_C$ by $R_{CD}$ around the minimum conductivity point exceeds the amount of $R_{CI}$, apparently negative $R_C$ values are observed.

## 4. DISCUSSION

In order to interpret the extracted $R_C$-$V_G$ characteristics shown in Fig. 3, it would be effective to examine the $V_G$ dependency of $R_{CI}$ and $R_{CD}$ individually.

Firstly, $R_{CI}$ is the actual contact resistance that relates to the carrier tunneling at metal-graphene interfaces, and the tunneling probability is proportional to the density of states (DOS) of the electrode metal at the Fermi level and that of graphene at the metal contacts.[5] Generally, $R_{CI}$ can be expressed as[15]

$$\frac{1}{W}\sqrt{\rho_{GC}\rho_C}\coth\left(\sqrt{\frac{\rho_{GC}}{\rho_C}}d_C\right), \qquad (1)$$

where $W$ is the contact width (the same as the channel width shown in Fig. 1a), $\rho_{GC}$ is the sheet resistance of graphene under the contact, $\rho_C$ is the contact resistivity, and $d_C$ is the contact length along the channel direction (1 μm in this study). Among these values, $W$ and $d_C$ are constants defined by the device structure. $\rho_C$ should originate from the carrier tunneling and is determined by the DOS, *i.e.*, the charge density of graphene under the metal contact. $\rho_{GC}$ is also determined by the charge density. However, the charge density of graphene under the contact is almost uncontrollable by $V_G$.[13,14] Therefore, charge-density pinning leads to the independence of $R_{CI}$ on $V_G$. If partial depinning is allowed at the contact, then $R_{CI}$ becomes $V_G$-dependent; in this case, the $R_{CI}$-$V_G$ characteristics are convex upward and cannot reproduce the dip around $V_G = V_{NP}$, as shown in Fig. 3 (see Appendix B for the modeling of $R_{CI}$).

Next, $R_{CD}$ can be estimated by empirical modeling of the graphene channel resistances employing $V_G$ dependency of the charge-density profile along the channel, which is shown in Fig. 4a. The details of the model have been reported previously.[9,10] Briefly, the channel resistance $R_G$ is obtained by integrating the local sheet resistance $\rho_G(x)$ along the channel, where $\rho_G(x)$ is uniquely determined by the local charge density shown in Fig. 4a. The parameters of the calculation are the length of the regions affected by the metal contacts $L_D$, the charge density at the metal contacts $n_D$, and the charge carrier mobility $\mu$ (see Appendix A for the modeling of $R_{CD}$). The thick solid line in Fig. 5a represents the $R_{CD}$-$V_G$ characteristics calculated for $L_D = 0.2$ μm, $n_D = 1.5 \times 10^{12}$ cm$^{-2}$ (the positive value indicates electron doping from metal contacts) and $\mu = 1.0$ m$^2$ V$^{-1}$ s$^{-1}$. The calculated characteristics are a good reproduction of the experimentally obtained dip around the minimum conductivity point. Consequently, if the depth of the dip is larger than the actual contact resistance $R_{CI}$, negative $R_C$ values are obtained by the TLM or four-terminal measurement.

Another important feature of the calculated $R_{CD}$-$V_G$ characteristics is the asymmetry between the positively and negatively gated regions. The dashed line in Fig. 5a shows the calculated $R_{CD}$-$V_G$ characteristics with the same conditions, except the polarity of the charge carriers doped from the metal contacts (*i.e.*, $L_D = 0.2$ μm, $n_D = -1.5 \times 10^{12}$ cm$^{-2}$ (hole doping) and $\mu = 1.0$ m$^2$ V$^{-1}$ s$^{-1}$). The overall shape of the characteristics is identical, except the asymmetry is opposite, *i.e.*, a mirror image. These results indicate that the carrier type doped from the metal contacts can be known by examining the asymmetry: namely, hole doping with higher $R_{CD}$ in the positively gated region, and electron doping with higher $R_{CD}$ in the negatively gated region. Thus, holes are doped from Ag and Cu contacts to graphene, and electrons are doped from Au contacts to graphene, which is the same as the carrier types deduced from the direction of the $V_{NP}$ shift accompanying the shortening of the channel lengths.[11]

The shape of the $R_C$-$V_G$ characteristics shown in Fig. 3 differs significantly for each metal. To clarify the origin of these differences, the $R_{CD}$-$V_G$ characteristics were calculated using variations of the $L_D$, $\mu$, and $n_D$ parameters, as shown in Figures 5b, 5c, and 5d, respectively. The fixed parameters were set to $L_D = 0.2$ μm, $\mu = 1.0$ m$^2$ V$^{-1}$ s$^{-1}$ and $n_D = 1.5 \times 10^{12}$ cm$^{-2}$ (electron doping). Figure 5b suggests that longer $L_D$ results in higher $R_{CD}$, but the overall shape of the $R_{CD}$-$V_G$ curves does not change with $L_D$. Figure 5c shows that the dip around $V_G = V_{NP}$ in the curves becomes less distinct (smaller and more broadened) with lower $\mu$. Figure 5d shows that there is almost no change in the sharpness of the dip for different $n_D$'s, although the asymmetric shape between the positively and negatively gated regions changes with $n_D$. Therefore, the experimentally obtained shapes for the $R_{CD}$-$V_G$ characteristics can be approximately fitted by determining the calculation parameters; (1) $\mu$ from the sharpness of the dip, (2) $n_D$ from examination of the curve shape, and (3) $L_D$ from the magnitude of the dip.

However, the experimental results also include the actual contact resistance $R_{CI}$. If a complete pinning of the charge density at the metal contacts is assumed, then $R_{CI}$ should not change with $V_G$. Thus, a constant $R_{CI}$ should be used to offset the baseline as the fourth parameter. The solid lines superimposed onto the experimental data shown in Fig. 3 are the best-fit lines using the four fitting parameters, and reproduce the dip structure well, although deviations are evident around the largest peak in Figs. 3b and 3c.[16] The extracted fitting parameters are compiled in Table I. Contacts that exhibit a higher charge density of graphene under the metal electrodes (higher $|n_D|$) resulted in a longer doping length $L_D$ and a lower actual contact resistance $R_{CI}$. Higher $|n_D|$ at the metal contacts increases the magnitude of the potential variation of graphene, which leads to deeper penetration of the metal-contact-induced potential variation into the graphene channel, *i.e.*, longer $L_D$.[17] In addition, $n_D$ regulates the DOS of graphene through the Fermi level shift, and thus controls the tunnel resistance at the metal-graphene interfaces. Higher $|n_D|$ indicates a higher DOS, which should result in a lower tunnel resistance $R_{CI}$.

The negative $R_C$ obtained for Ag contacts indicates that $|R_{CD}| > R_{CI}$ is achieved around the minimum conductivity point at the Ag-graphene interface. This condition can be fulfilled if the effect of metal-contact doping is substantial, while charge injection from the metal contacts is efficient, as shown in Fig. 6. At the interfaces of metal electrodes and typical semiconductors, the potential variation induced by the metal contacts can be significantly large, but the metal-contact doping effect should be limited due to the finite bandgap of the semiconductor (low $|R_{CD}|$). On the other hand, the carrier injection is considered to be an inefficient thermionic emission process (high $R_{CI}$), which should result in $|R_{CD}| < R_{CI}$. At metal-metal interfaces, the carrier injection should be much more efficient than that at the metal-semiconductor interfaces (low $R_{CI}$); however, electrostatic screening inhibits the potential variation inside the metal (very low $|R_{CD}|$), which should also lead to $|R_{CD}| < R_{CI}$. At the metal-graphene interfaces, no bandgap in graphene ensures relatively efficient charge injection by a tunneling process (low $R_{CI}$). The strongly suppressed screening in graphene[17] induces a long-range potential variation, and the no bandgap of graphene allows the formation of highly doped regions near the metal contacts (high $|R_{CD}|$), which can fulfill $|R_{CD}| > R_{CI}$ in some cases. The considerations for graphene should be applicable to other systems that have similar electronic structures, *i.e.*, Dirac-cone systems.

## 4. CONCLUSIONS

The gate-voltage dependency of the contact resistances in graphene FETs have been characterized in detail. The experimental $R_C$-$V_G$ characteristics obtained using the TLM for Ag, Cu, and Au contacts exhibit a dip around the minimum conductivity point ($V_G = V_{NP}$), and $R_C$ at this point became negative for Ag contacts. $R_C$ extracted by the TLM or four-terminal measurement should include a contribution from an additional resistance due to metal-contact doping $R_{CD}$, in addition to the actual tunnel resistance precisely at the metal-graphene contacts, $R_{CI}$. A diffusive-transport model, which includes charge carrier doping from metal electrodes to graphene channels and charge-density pinning at the metal contacts, was developed to examine the $R_{CD}$-$V_G$ characteristics. The model was found to reproduce the experimentally observed dip structure well using the TLM. The apparently negative $R_C$ observed with Ag contacts is considered to be a characteristic feature of Dirac-cone systems such as graphene, and $R_C$ at the interfaces of metals with other systems (metals or typical semiconductors) should be positive.

**ACKNOWLEDGEMENTS**

The authors acknowledge M. Murakami of Kaneka Corporation and M. Shiraishi of Osaka University for providing the graphite crystal. This work was supported by grants (Kakenhi; Grant Nos. 20710071 and 19051001) and World Premier International Research Center Initiative (WPI) from the Ministry of Education, Culture, Sports, Science and Technology (MEXT), Japan.

**APPENDIX A: MODELING OF $R_{CD}$**

An empirical model for the channel resistance of graphene FETs was developed by taking into consideration the experimental features such as the minimum conductivity, metal-contact doping, and charge-density pinning.[9,10] In the model, the channel resistance $R_G$ is expressed as

$$R_{\mathrm{G}} = \frac{1}{W}\int_0^L \rho_{\mathrm{G}}(x)dx = \frac{1}{W}\int_0^L \left[\left(\mu\frac{\varepsilon_0\varepsilon_{\mathrm{r}}}{d}V(x)\right)^2 + (\sigma_{\min})^2\right]^{-\frac{1}{2}} dx \quad , \qquad (A.1)$$

where $\varepsilon_0$ is the permittivity of a vacuum, $\varepsilon_{\mathrm{r}}$ is the relative permittivity, $d$ is the thickness of the gate dielectric, $V(x)$ is the $V_{\mathrm{G}}$ equivalent corresponding to the local carrier density at position $x$, and $\sigma_{\min}$ is the minimum conductivity. $V(x)$ can be obtained by determining the $V_{\mathrm{G}}$ dependence of the charge-density profile along the channel, which is shown in Fig. 4a, as

$$V(x) = -\frac{V_{\mathrm{D}}-V_{\mathrm{G}}}{L_{\mathrm{D}}}x + V_{\mathrm{d}} \quad (0 \le x \le L_{\mathrm{D}}),$$
$$V(x) = V_{\mathrm{G}} \quad (L_{\mathrm{D}} \le x \le L - L_{\mathrm{D}}), \qquad (A.2)$$

where $V_{\mathrm{D}}$ represents the doping level at the contact edges ($= V(0) = V(L)$). By substituting Eq. (A.2) into Eq. (A.1), $R_{\mathrm{G}}$ for the completely-pinned contacts becomes[10]

$$R_{\mathrm{G}} = \frac{L-2L_{\mathrm{D}}}{W}\left[\left(\mu\frac{\varepsilon_0\varepsilon_{\mathrm{r}}}{d}V_{\mathrm{G}}\right)^2 + (\sigma_{\min})^2\right]^{-\frac{1}{2}} + R_{\mathrm{D}},$$

$$R_{\mathrm{D}} = \frac{2L_{\mathrm{D}}}{W}\left[\left(\mu\frac{\varepsilon_0\varepsilon_{\mathrm{r}}}{d}V_{\mathrm{D}}\right)^2 + (\sigma_{\min})^2\right]^{-\frac{1}{2}} \quad (\text{for } V_{\mathrm{G}} = V_{\mathrm{D}}),$$

$$R_{\mathrm{D}} = \frac{-2L_{\mathrm{D}}}{W}\frac{1}{\mu}\frac{d}{\varepsilon_0\varepsilon_{\mathrm{r}}}\frac{1}{V_{\mathrm{D}}-V_{\mathrm{G}}}\left\{\ln\left|V_{\mathrm{G}}+\sqrt{V_{\mathrm{G}}^2+\left(\frac{1}{\mu}\frac{d}{\varepsilon_0\varepsilon_{\mathrm{r}}}\sigma_{\min}\right)^2}\right| - \ln\left|V_{\mathrm{D}}+\sqrt{V_{\mathrm{D}}^2+\left(\frac{1}{\mu}\frac{d}{\varepsilon_0\varepsilon_{\mathrm{r}}}\sigma_{\min}\right)^2}\right|\right\} \quad (\text{for } V_{\mathrm{G}} \ne V_{\mathrm{D}}).$$
$$(A.3)$$

Using these developed expressions, the "contact" resistance which originates from the channel region $R_{\mathrm{CD}}$ can be easily calculated by

$$R_{\mathrm{CD}} = \frac{1}{2}[R_{\mathrm{G}}]_{L=0} = -\frac{2L_{\mathrm{D}}}{W}\left[\left(\mu\frac{\varepsilon_0\varepsilon_{\mathrm{r}}}{d}V_{\mathrm{G}}\right)^2 + (\sigma_{\min})^2\right]^{-\frac{1}{2}} + R_{\mathrm{D}} \quad . \qquad (A.4)$$

It should be noted that $R_{\mathrm{CD}}$ is not the actual contact resistance, but it becomes apparent due to the methodologies used for contact-resistance extraction by the TLM or four-terminal measurement. All simulation results were calculated for graphene on a common $SiO_2$ gate dielectric ($\varepsilon_{\mathrm{r}} = 3.9$, $d = 300$ nm), and $\sigma_{\min} = 4e^2/h$ is employed,[18] where $e$ is the elementary charge and $h$ is Planck's constant. $n_{\mathrm{D}}$ introduced in the main text relates to $V_{\mathrm{D}}$ as

$$n_\mathrm{D} e = C_0 V_\mathrm{D} = \frac{\varepsilon_0 \varepsilon_\mathrm{r}}{d} V_\mathrm{D}, \tag{A.5}$$

where $C_0$ is the capacitance of the gate dielectric per unit area.

**APPENDIX B: MODELING OF $R_\mathrm{CI}$**

As described in the main text, the actual contact resistance $R_\mathrm{CI}$ is generally expressed as[15]

$$\frac{1}{W} \sqrt{\rho_\mathrm{GC} \rho_\mathrm{C}} \coth\left(\sqrt{\frac{\rho_\mathrm{GC}}{\rho_\mathrm{C}}} d_\mathrm{C}\right). \tag{B.1}$$

$\rho_\mathrm{GC}$ is the sheet resistance of graphene under the contact; from Eq. (A.1), it can be written as

$$\rho_\mathrm{GC} = \left[\left(\mu \frac{\varepsilon_0 \varepsilon_\mathrm{r}}{d} V_\mathrm{D}\right)^2 + (\sigma_\mathrm{min})^2\right]^{-1/2}. \tag{B.2}$$

$\rho_\mathrm{C}$ is the contact resistivity, which should be related to the carrier tunneling at metal-graphene interfaces, and the tunneling probability is proportional to the DOS of the electrode metal at the Fermi level and that of graphene at the metal contacts.[5] Provided that the DOS of the electrode is constant, then $\rho_\mathrm{C}$ becomes inversely proportional to the DOS of graphene, *i.e.*, to the absolute energy difference $|\Delta E|$ between the Fermi level and the Dirac point.[19] $|\Delta E|$ corresponds to $n_\mathrm{D}$ according to[20,21]

$$\Delta E = \mathrm{sgn}(n_\mathrm{D}) \hbar v_\mathrm{F} \sqrt{\pi |n_\mathrm{D}|}, \tag{B.3}$$

where $\hbar$ is the Dirac constant and $v_\mathrm{F}$ is the Fermi velocity (ca. $10^6$ m s$^{-1}$). Thus, $\rho_\mathrm{C}$ can be expressed as

$$\rho_\mathrm{C} \propto \frac{1}{|\Delta E|} \propto \frac{1}{\hbar v_\mathrm{F} \sqrt{\pi |n_\mathrm{D} e|}} = \frac{1}{\hbar v_\mathrm{F} \sqrt{\pi C_0 |V_\mathrm{D}|}} \propto |V_\mathrm{D}|^{-1/2}. \tag{B.4}$$

Complete pinning of the charge density of graphene under the contact leads to constant $\rho_\mathrm{GC}$ and $\rho_\mathrm{C}$, and thus constant $R_\mathrm{CI}$ with respect to $V_\mathrm{G}$. If partial depinning is allowed, then the charge density of graphene under the metal contacts should be determined by $V_\mathrm{D} + pV_\mathrm{G}$ instead of $V_D$ by introducing the depinning factor $p$ (indicative of the degree of depinning), defined as 0 and 1 for pinned and depinned contacts, respectively. For simplicity, $R_{CI}$ is examined by separating Eq. (B.1) into two parts:

$$\frac{1}{W}\sqrt{\rho_{GC}\rho_C} \propto \sqrt{\frac{1}{\sqrt{\left[\mu\frac{\varepsilon_0\varepsilon_r}{d}(V_D+pV_G)\right]^2+(\sigma_{min})^2}}\frac{1}{\sqrt{|V_D+pV_G|}}},$$

$$\coth\left(\sqrt{\frac{\rho_{GC}}{\rho_C}}d_C\right) = \coth\left(A\sqrt{\frac{\sqrt{|V_D+pV_G|}}{\sqrt{\left[\mu\frac{\varepsilon_0\varepsilon_r}{d}(V_D+pV_G)\right]^2+(\sigma_{min})^2}}}d_C\right),$$

(B.5)

where $A$ is a proportionality constant. Figure 7 shows the simulated $V_G$ dependency of these two parts of $R_{CI}$, which were calculated using $\mu = 0.5$ m$^2$ V$^{-1}$ s$^{-1}$ and $V_D = 20$ V (electron concentration of $n_D = 1.4\times10^{12}$ cm$^{-2}$) with $p = 0$, 0.2, 0.5, and 1. Both curves exhibit a divergence at $V_G = -V_D/p$, where the charge density of graphene under the metal contacts is equal to zero. In actual devices, graphene has inhomogeneity, and thus a charge density of zero cannot be achieved; the finite DOS at electron/hole puddles[22] avoids the divergence, so that a peak with a finite height is observed instead.

FIGURES:

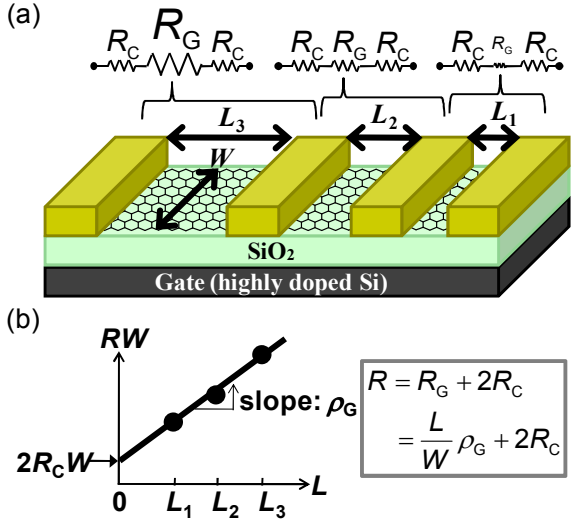

FIG. 1. (Color online) (a) Schematic of the fabricated device structure with several electrodes and different inter-electrode spacings $L$, on a single flake of graphene. (b) Schematic of the TLM. A contact resistance can be extracted from an intercept at $L = 0$ for a linear fitting of the two-terminal resistance, $R$, versus $L$ plot.

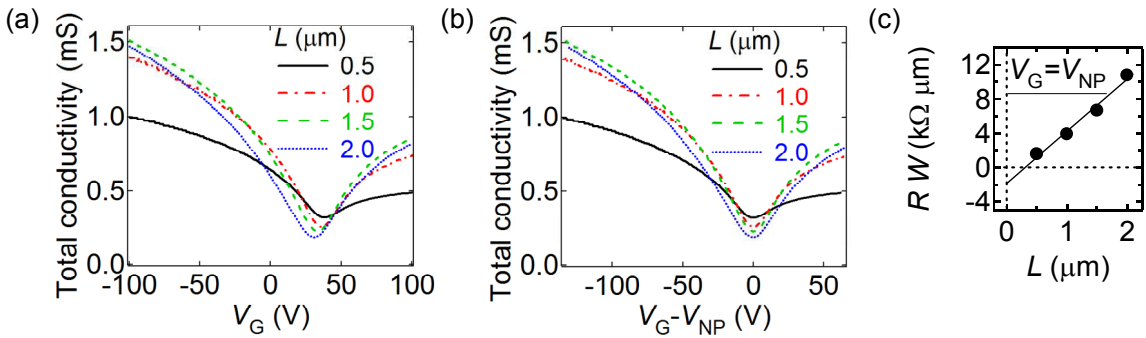

FIG. 2. (Color online) (a,b) Transfer characteristics of Ag-contacted graphene FETs with different channel lengths ($L = 2.5$ to $0.5$ μm) fabricated on the same graphene flake. The characteristics in (b) are replots of (a) against the normalized gate voltage, $V_G − V_{NP}$. (c) $R$-$L$ plot of the Ag-contacted devices at $V_G = V_{NP}$. The straight line is the result of a least-squares fit, which indicates an apparently negative contact resistance.

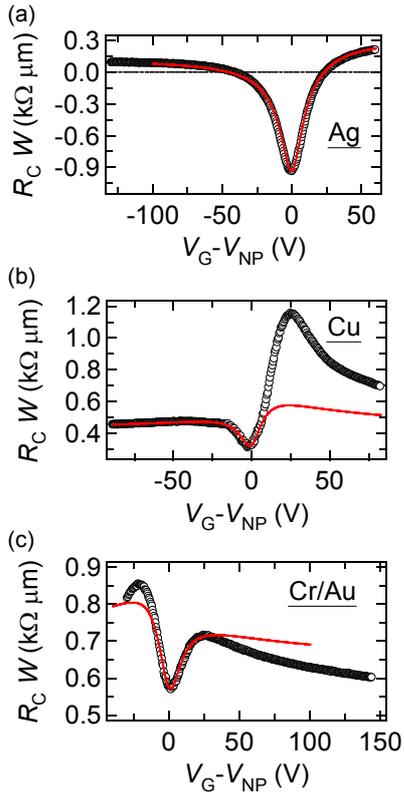

FIG. 3. (Color online) Dependence of the contact resistance on the gate voltage extracted using the TLM for (a) Ag, (b) Cu, and (c) Au contacts. The solid lines represent the best-fit lines of calculations using the proposed model. Deviation of the fitting lines from the experimental curves may be due to partial depinning of the metal contacts.[16]

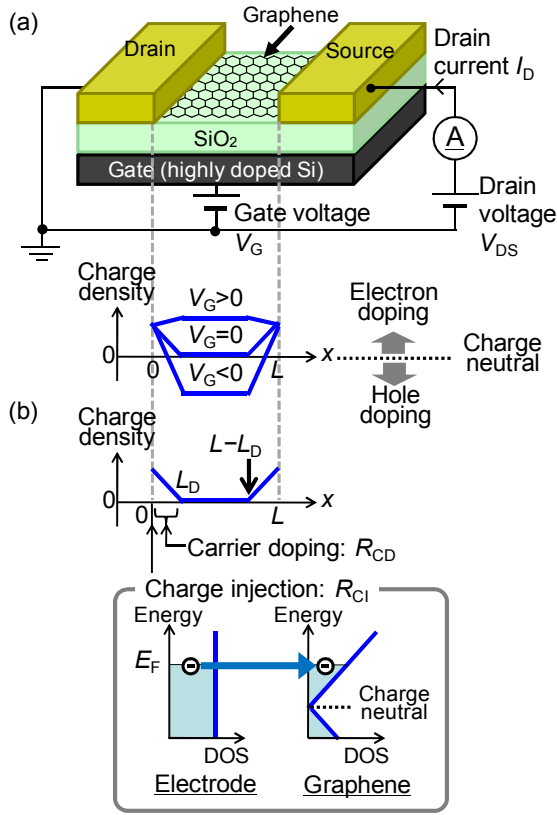

FIG. 4. (Color online) (a) Gate-voltage dependence of the charge-density profile for the model calculation. The vertical axis is a gate-voltage equivalent, and positive (negative) values correspond to electron (hole) doping. Linearly graded doping occurs from the metal contact edges ($x = 0, L$) to the distance into the graphene channel, $L_D$. The charge density at the metal contacts is uncontrollable with the application of gate voltages. (b) Two possible contributions to the contact resistances extracted by the TLM or four-terminal measurement are 1) the actual contact resistance precisely at the contact, $R_{CI}$ (most simply, ascribable to a tunneling process[5]), and 2) the apparently-appeared contact resistance due to the potential variation induced by the metal contact, $R_{CD}$.

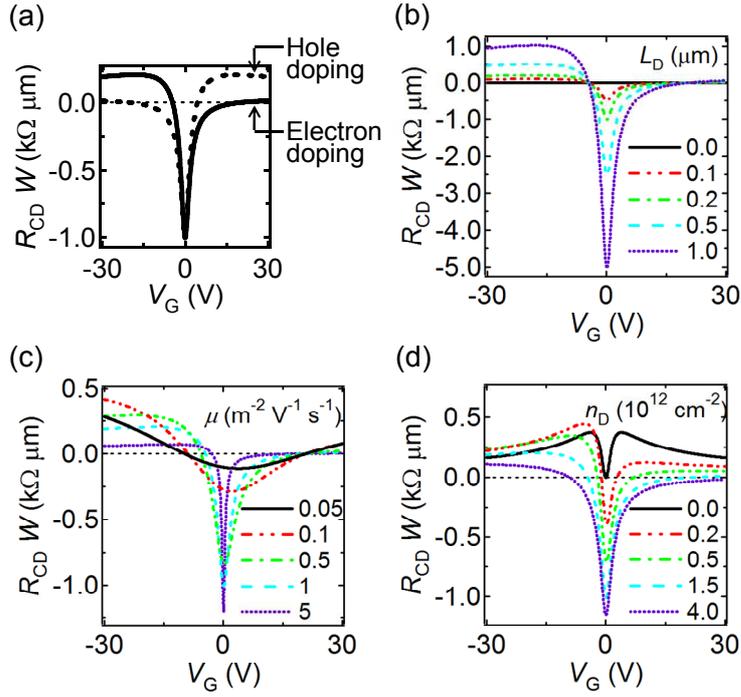

FIG. 5. (Color online) Simulated $R_{CD}$-$V_G$ characteristics with variation of the (a) polarity of doped carriers, (b) $L_D$, (c) $\mu$, and (d) $n_D$. The fixed parameters were set to $L_D = 0.2$ μm, $\mu = 1.0$ m$^2$ V$^{-1}$ s$^{-1}$ and $n_D = 1.5\times 10^{12}$ cm$^{-2}$. The solid and dashed lines in (a) indicate the results for positive (electron doping) and negative (hole doping) $n_D$, respectively.

| material | $R_{CI}$ | $R_{CD}$ | comparison |
|---|---|---|---|
| semi-conductor | **High** — Thermionic | **Low** — Finite bandgap & weak screening | $|R_{CD}| < R_{CI}$ |
| metal | **Low** — Tunneling | **Very low** — Strong screening | $|R_{CD}| < R_{CI}$ |
| Dirac-cone system | **Low** — Tunneling | **High** — No bandgap & weak screening | $|R_{CD}| > R_{CI}$ |

FIG. 6. (Color online) Comparison of the magnitudes of $R_{CI}$ and $R_{CD}$ at metal contacts to various systems. The condition $|R_{CD}| > R_{CI}$, which is necessary to observe the apparently negative $R_C$ as extracted in this study, is considered to be fulfilled at the interfaces of metals with Dirac-cone systems.

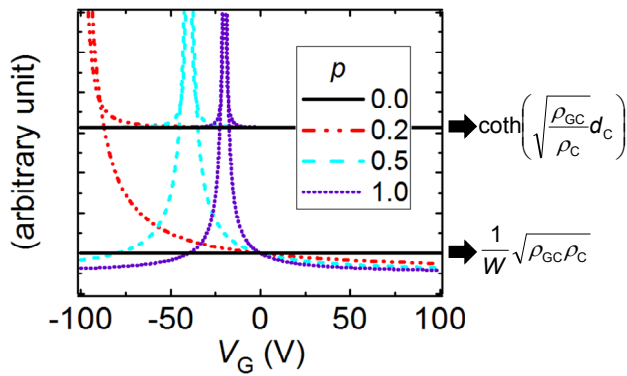

FIG. 7. (Color online) Simulated $V_G$ dependency of the two parts of $R_{CI}$ with various $p$. While the completely pinned contacts ($p = 0$) lead to no $V_G$ dependence, the fully or partially depinned contacts (finite $p$) induce a divergence at $V_G = -V_D/p$.

Table:

TABLE I. $R_C$ simulation parameters extracted by fitting to the experimentally obtained $R_C$-$V_G$ characteristics shown in Figure 3.

| contact metal | $\mu$ (m$^2$ V$^{-1}$ s$^{-1}$) | $n_D$ (10$^{12}$ cm$^{-2}$) | $L_D$ (μm) | $R_{CI} \cdot W$ (kΩ μm) |
|---|---|---|---|---|
| Ag | 0.18 | −11.5 | 0.20 | 0.10 |
| Cu | 0.18 | −0.6 | 0.08 | 0.39 |
| Au | 0.16 | 0.5 | 0.08 | 0.62 |